# Plots for visualizing paper impact and journal impact of single researchers in a single graph

Lutz Bornmann* & Robin Haunschild**

* Corresponding author:

Division for Science and Innovation Studies

Administrative Headquarters of the Max Planck Society

Hofgartenstr. 8,

80539 Munich, Germany.

E-mail: bornmann@gv.mpg.de

** Max Planck Institute for Solid State Research,

Heisenbergstr. 1,

70569 Stuttgart, Germany.

Email: R.Haunschild@fkf.mpg.de

**Abstract**

In research evaluation of single researchers, the assessment of paper and journal impact is of interest. High journal impact reflects the ability of researchers to convince strict reviewers, and high paper impact reflects the usefulness of papers for future research. In many bibliometric studies, metrics for journal and paper impact are separately presented. In this paper, we introduce two graph types, which combine both metrics in a single graph. The graphs can be used in research evaluation to visualize the performance of single researchers comprehensively.

# 1 Introduction

Publication success of single researchers can be assessed bibliometrically on the paper and journal basis. Both perspectives provide different insights. Researchers submit their manuscripts to journals and these journals can be high-impact or low-impact journals. Thus, it is a first success of researchers if they are able to get a high share of their manuscripts accepted in high-impact journals. For the publication in high impact journals, very skeptical reviewers and editors have to be convinced (Bornmann, 2011). High-impact journals can be multi-disciplinary journals, like *Science* or *Nature*, or subject-specific journals, like *Angewandte Chemie – International Edition*. Since the impact of the journals depends on their reputation (besides other factors), the ability to publish papers in high-impact journals reflects one aspect of high performance. Several years subsequent to the publication of a manuscript in a journal (at least three years), the citation impact of the published manuscript can be determined. Citations are seen as a proxy of quality, which reflects at least impact (but not, e.g., accuracy). Thus, high citations for the researcher's papers indicate his or her ability to publish influential research. This is a second success, which can be reached by researchers.

In research evaluation of single researchers, the assessment of both events is interesting. High journal impact reflects the ability to convince strict reviewers and editors, and high paper impact reflects the usefulness of papers for future research (Bornmann & Marx, 2014c). In many bibliometric studies, metrics for journal and paper impact are separately presented. In this paper, we introduce two graph types, which combine both metrics in a single graph. These plots analyze the magnitude of differences between two metrics (here: journal and paper impact). The proposed graph types are able to show the share of a researcher's papers with higher paper impact than can be expected from the journal impact.

The proposals in this paper follow earlier proposals of visualizing bibliometric data of single researchers with beamplots (Bornmann & Marx, 2014a, 2014b). Beamplots show the performance of single researchers in view of publication output and citation impact.

## 2 Methods

Percentiles have been introduced in bibliometrics to overcome a certain problem with frequently used citation indicators: the indicators are based on the arithmetic average. However, since, as a rule, the distribution of citations across publications is skewed, the arithmetic average should not be used with raw citation data. The citation percentile of a single paper is an impact value below which a certain share of papers fall (Bornmann, Leydesdorff, & Mutz, 2013). Given a set with several papers and their citation counts, the citation percentile of 95 for a focal paper means, for example, that 95% of the papers in a set have citations counts lower than the focal paper. The formula by Hazen (1914) ($(i - 0.5) / n * 100$) is frequently used for the calculation of percentiles, whereby $n$ is the number of papers in the set and $i$ is the rank position of the papers in the set (concerning their citations). In bibliometrics, the set of papers is defined field- and time-specific (e.g., by using Web of Science subject categories and publication years). The corresponding citation percentile for each paper is a field- and time-normalized impact value, which can be used for cross-field comparisons.

In this study, we used citation percentiles, which are based on Hazen's formula, for the papers of two single anonymous researchers as examples to demonstrate the plots. Researcher 1 has published 99 papers (articles and reviews) between 2004 and 2013; researcher 2 has published 427 papers between 1997 and 2013. Citations were counted until the end of 2016 for calculation of Hazen percentiles. For field classification, we used the Web of Science (WoS, provided by Clarivate Analytics, formerly the IP and Science business of Thomson Reuters) subject categories. In the case of multiple assignments of papers to subject categories, the average value of the percentile in each subject category was used to obtain a paper-based percentile value (Haunschild & Bornmann, 2016). We retrieved Hazen percentiles for the papers from our in-house database derived from the Science Citation Index Expanded (SCI-E), Social Sciences Citation Index (SSCI), and Arts and Humanities Citation Index (AHCI) provided by Clarivate Analytics. Percentiles on the paper basis are not only available in our in-house database (or similar databases at other bibliometric institutes), but also in advanced bibliometric tools, such as InCites (Clarivate Analytics) and SciVal (Elsevier).

Table 1 shows some metrics for both researchers. These metrics have been recommended by Bornmann and Marx (2014b) for the evaluation of single researchers. Highly-cited papers are those papers, which belong to the 10% most-frequently cited papers in the corresponding subject categories and publication years. For the calculation of the age-normalized number of highly-cited papers, the number of highly-cited papers is divided by the number of years since publishing the first paper. This metric has been favored by Bornmann and Marx (2014b) against the use of the h index (Hirsch, 2005), because it is a field- and age-normalized index. A further advantage of the age-normalized number of highly-cited papers is that it does not work with the arbitrary threshold h to identify the papers with the most impact. It uses for each researcher the same threshold: being among the 10% most-frequently cited.

Table 1. Metrics for two researchers

|  | Researcher 1 | Researcher 2 |
|---|---|---|
| Number of papers | 99 | 427 |
| Number of highly-cited papers | 29 | 254 |
| Proportion of highly-cited papers | 29% | 60% |
| Age-normalized number of highly-cited papers | 3 | 15 |

Recently, Pudovkin and Garfield (2004) introduced a similar metric as citation percentiles on the level of journals. Also, this metric leads to field-normalized values – on the level of journals. The so-called rank normalized impact factor (rnIF) equals (($k - r_j + 1$) / $k$ *100) where $r_j$ is the descending rank of journal $j$ in its subject category and $k$ is the number of journals in the category. In contrast to the usual Journal Impact Factor (Garfield, 2006), the rnIF can be used for cross-field comparisons of journals. Our in-house database also provides the journal percentiles (rnIF) from the Journal Citation Reports (Clarivate Analytics). The corresponding journal percentiles for the papers under study were retrieved.

In the following, we call citation percentiles on the level of papers as paper percentiles and citation percentiles on the level of journals (rnIF) as journal percentiles. By using both percentiles for a publication set of a single researcher, we have similar field-normalized metrics available, which can be used for the comparison of two events: the success of publishing in good journals and the success of receiving high citation counts. If average percentiles are reported in the following, these averages are medians.

We provide a Stata command (babibplot.ado) and an R package (BibPlots), which produce the graphs proposed in this study. Both can be found in SSC Archive (in the case of Stata) and CRAN (Comprehensive R Archive Network, in the case of R), respectively.

## 3    Results

Bornmann and Marx (2014a, 2014b) introduced beamplots, which can be used to visualize the productivity and citation impact of single researchers. It is an advantage of the plots that they contain distributional information (the spread of papers across citation percentiles and publication years) and index information (the mean impact of the papers over all publication years and the mean impact within single publication years). Bornmann and Marx (2014a, 2014b) propose to use beamplots with paper percentiles, but they can also be used with journal percentiles.

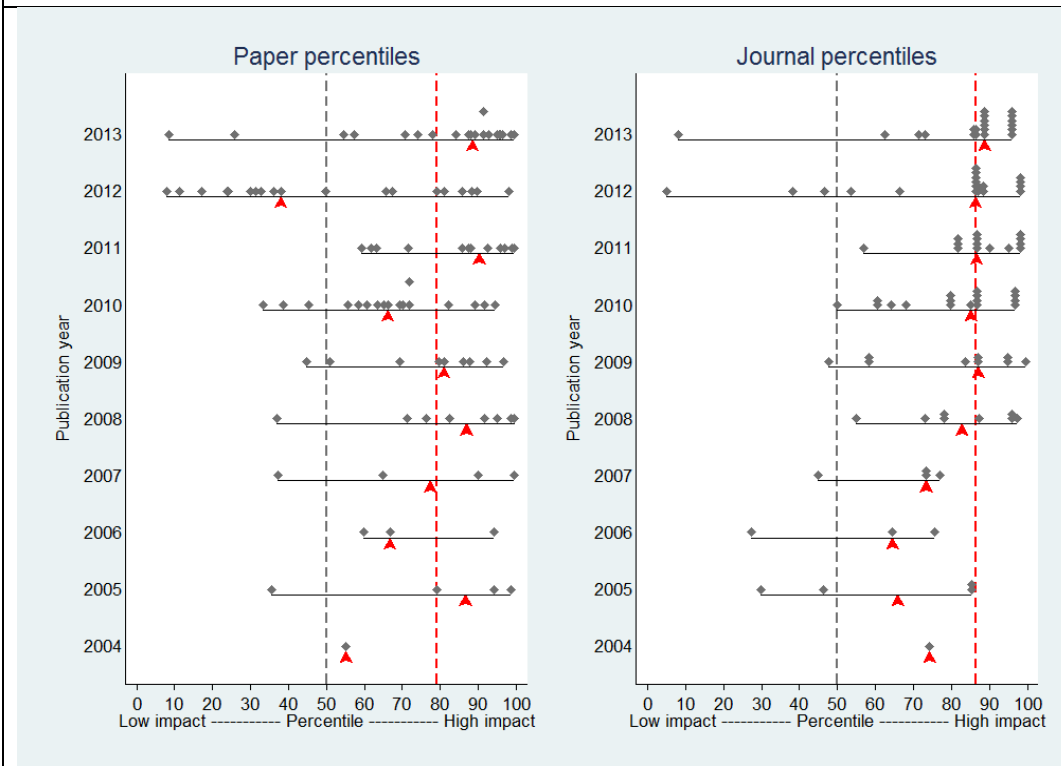

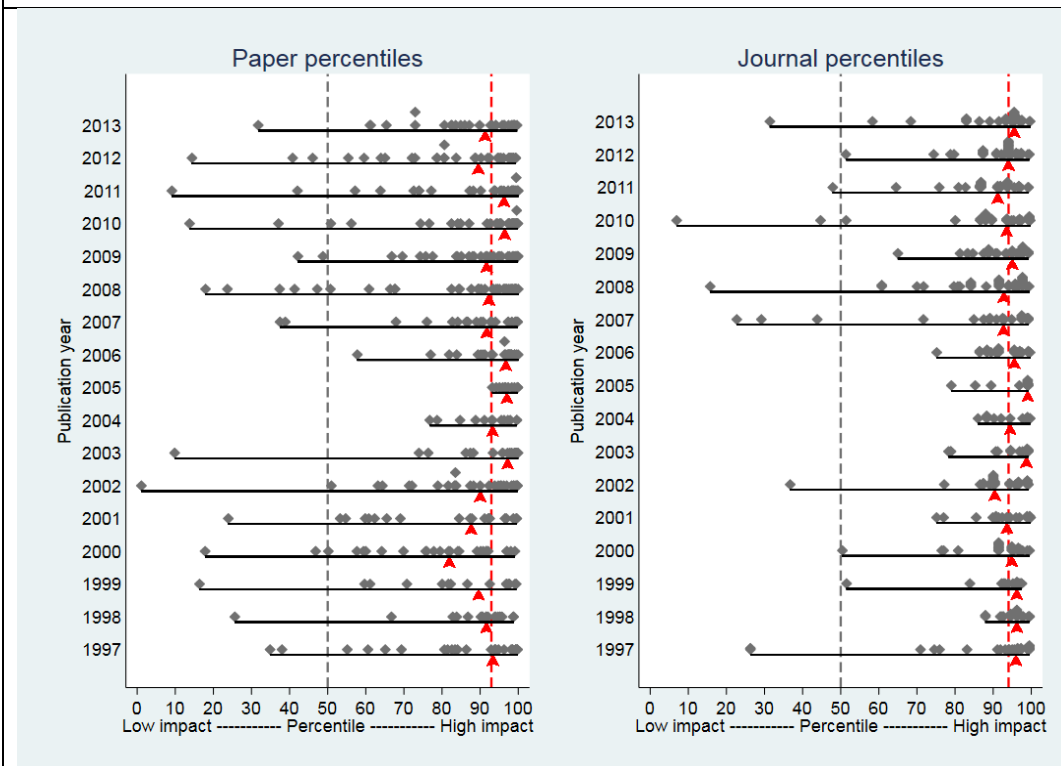

Figure 1. Beamplots of paper and journal percentiles

Figure 1 presents beamplots of paper (left) and journal (right) percentiles for both researchers. The individual percentiles (paper or journal percentiles) are shown using grey rhombi; the median over a publication year is displayed with red triangles. Furthermore, a red dashed line visualizes the median of the (paper or journal) percentiles for all the years and a grey line marks the value 50 (the median). Whereas the metrics in Table 1 only condense the outcome of a career in only a few numbers, beamplots provide output and input information for every year. Thus, it can be inspected, in which periods the researcher was very productive and successful or not. Figure 1 demonstrates, for example, that researcher 1 has considerable variations across the career, but this is scarcely visible for researcher 2.

Although the annual ability to publish high impact papers and papers in high impact journals can be seen in the beamplots of both researchers, the connection between the two is lost for the years where multiple papers with different paper impact were published. Thus, we propose the possibility to keep the connection between paper and journal percentiles for each data point in a scatter plot, as shown in Figure 2. The horizontal and vertical red lines in the figure indicate the world averages; the red dashed lines show the average values of paper and journal percentiles in the data set. The diagonal red line is the bisecting line. Points below the bisecting line indicate that the corresponding paper has a higher paper impact than journal impact and vice versa. Each row ($n_r$) and column ($n_c$) as well as quadrant ($n_q$) of the scatter plot is labeled with the number and percentage of data points in the corresponding section, e.g., $n_{r1}$ = 89; 90% for 89 papers (90%) in row 1 of researcher 1. The values of $n_{c1}$ correspond to the number and proportion of papers belonging to the 50% most frequently cited papers in the corresponding subject categories and publication years. The red squares show the average value of all data points in each quadrant. The results in Figure 2 demonstrate that both researchers were able to publish most of the papers in better than average journals with better than average impact later on. This is especially the case for researcher 2.

Researcher 1

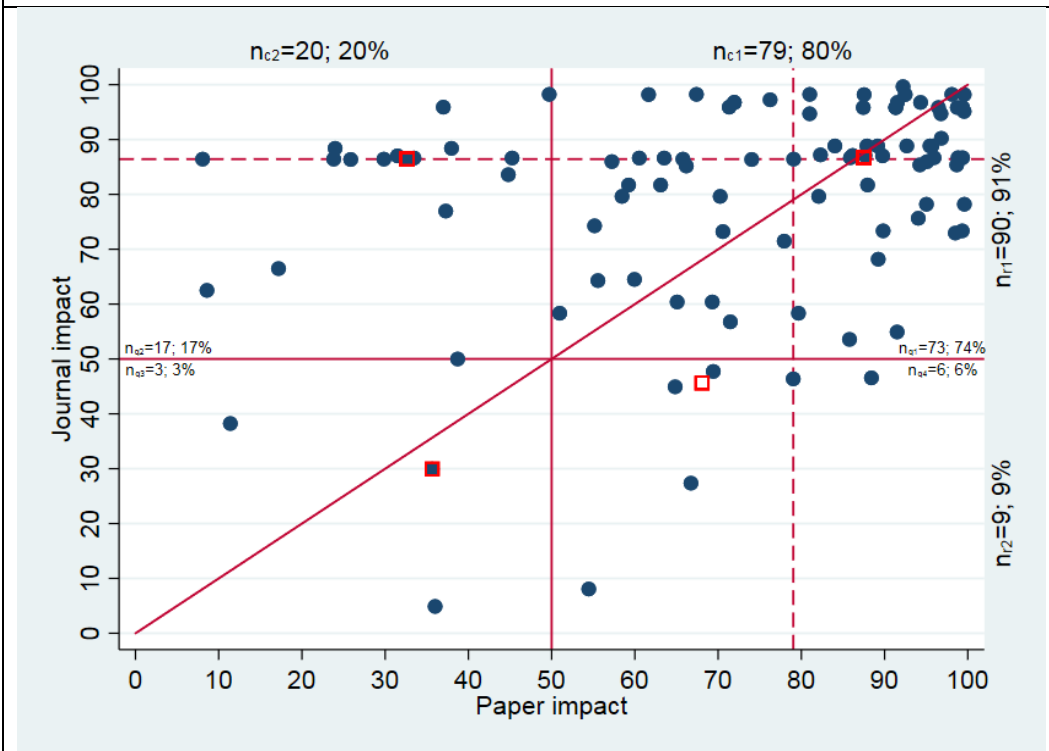

Researcher 2

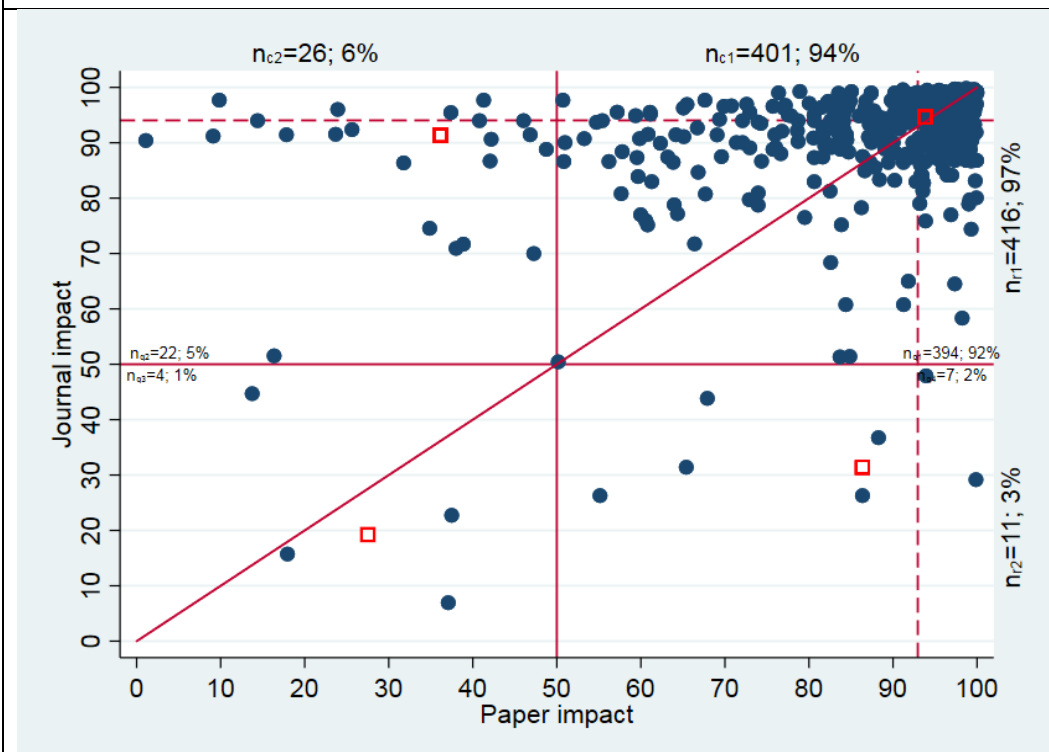

Figure 2. Scatter plots of paper and journal percentiles (both variables are interdependent and the axes can be switched)

Researcher 1

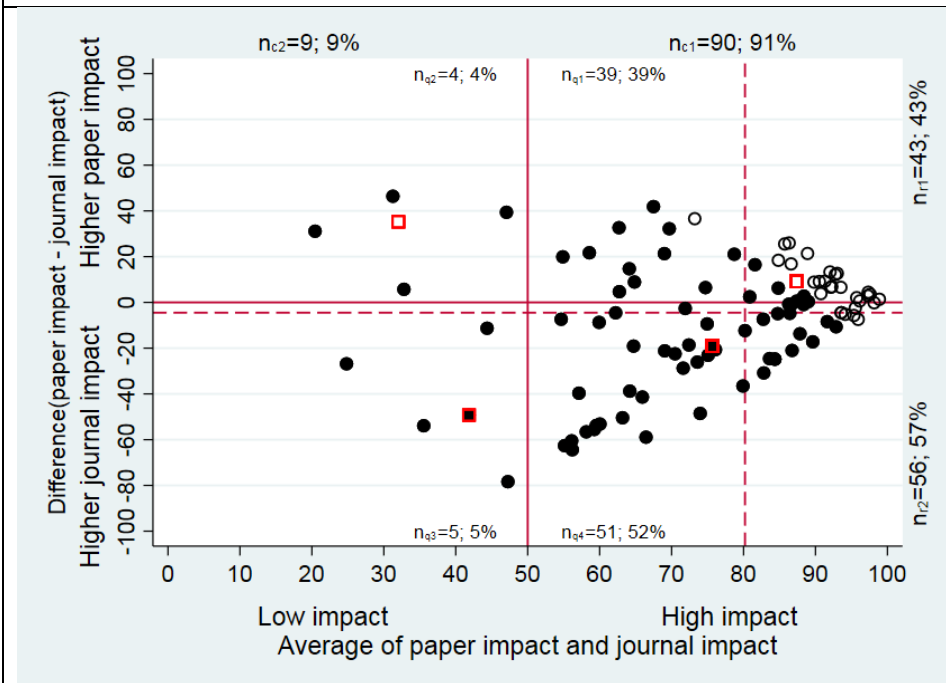

Researcher 2

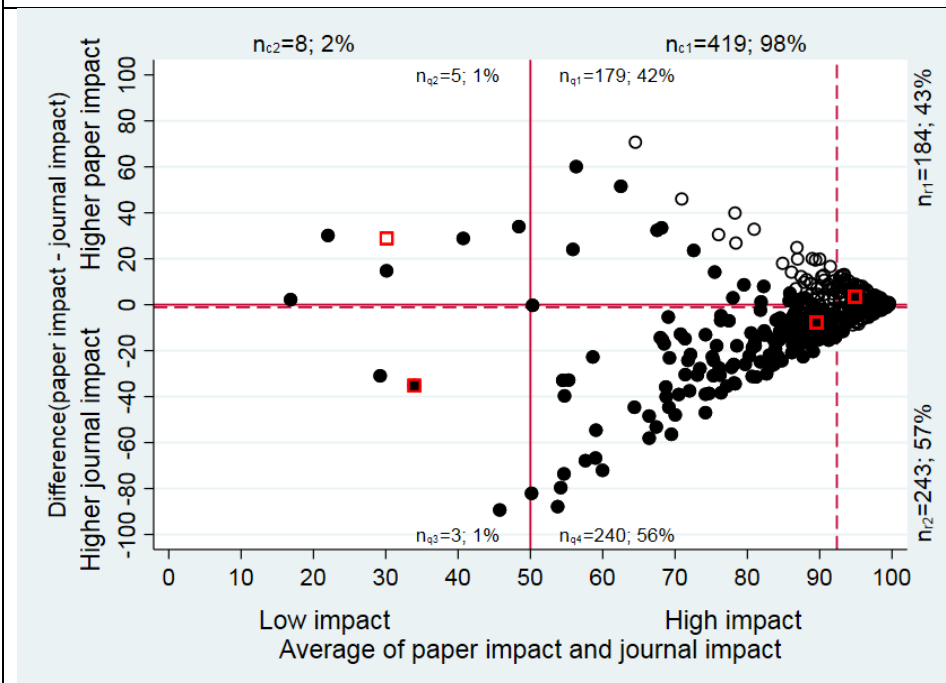

Figure 3. Difference against mean plots of paper and journal percentiles. Unfilled circles are papers, which belong to the 10% most frequently cited papers in the corresponding subject categories and publication years; the papers with less impact are filled circles (both variables are interdependent and the axes can be switched)

Figure 3 emphasizes the relation between paper impact and journal impact even more. The figures for the researchers are difference against mean plots (Altman & Bland, 1983; Cleveland, 1985); they are scatter plots of the following two quantities: (1) difference between paper and journal impact and (2) average of paper and journal impact. The bisecting line is no longer necessary to judge whether the paper percentile is higher than the journal percentile or vice versa. Additionally, the plots in Figure 3 feature two dashed red lines, which answer the following questions: (1) is there a general tendency of the researcher to publish in journals with higher impact or to publish papers with higher impact (see the y-line)? (2) Is the researcher generally able to publish papers in good journals, which receive high impact later on (see the x-line)? By combining both quantities as two new indexes in the same plot, we provide additional information compared to journal and paper percentiles. Furthermore, the individual paper and journal impact values remain easily accessible: (1) paper impact = half of the difference + average and (2) journal impact = average - half of the difference.

Often, one is interested in identifying the papers which belong to the 10% most frequently cited papers in the corresponding subject categories and publication years (Bornmann, de Moya Anegón, & Leydesdorff, 2012). This is visible in the scatter plots in Figure 2. The data points with a paper impact of 90 or higher are among the top 10% of their subject category and in their publication year. However, this information is lost in the difference against mean plots of Figure 3. In order to restore this information in the plots, the black circles with a paper impact of 90 or higher (papers among the top 10%) are unfilled while the other circles are filled.

A benefit of the difference against mean plot lies in the analysis of the relationship between the differences of journal and paper impact and their average. Papers with extreme differences between journal and paper impact are clearly identifiable. The difference against mean plots in Figure 3 combines the need to see the distribution of the impact of the individual papers and the journals they are published in with aggregated statistical values over the quadrants, rows, and columns of the plot. Furthermore, the average of the differences between paper and journal impact (see the horizontal dashed line in the plots of Figure 3) shows whether there is a general tendency of the researchers to publish in better journals or papers with higher impact. The vertical dashed lines in the plots of Figure 3 show the overall average impact (paper and journal impact).

## 4  Discussion

Researchers and decision makers in science policy are usually interested in receiving single numbers on the performance of single researchers (Leydesdorff, Wouters, & Bornmann, 2016). This explains the popularity of the h index proposed by Hirsch (2005) and the many variants of this index introduced in recent years (Bornmann, Mutz, Hug, & Daniel, 2011). However, since the reduction of the performance into a single number leads to the loss of information, it is recommended in bibliometrics to use distributions instead of only single numbers. For example, Lariviere et al. (2016) propose to use a method for generating the citation distribution of journals, instead of the use of the Journal Impact Factor (JIF) in research evaluation. Basically, the JIF measures the mean citation impact of papers published in a journal. Similarly, the performance of single scientists should be also measured by focusing on distributions than only on single numbers. In this study, we have demonstrated the usefulness of scatter plots and difference against mean plots to combine paper and journal percentiles in a single graph. The availability of journal percentiles in the Journal Citation Reports (JCR, Clarivate Analytics) makes it possible to contrast paper impact with journal impact – time- and field-normalized by using percentiles.

Difference against mean plots are a standard instrument in the assessment of equivalence between two metrics (usually two clinical measurement methods). We think that the plots also allow interesting insights into the publication and citation profiles of single scientists and can be used for research evaluation purposes. The plots can be applied to inspect the amount of papers in a set, which – more or less – agree in a high or low paper and journal impact. This is the average dimension of the plot. The other dimension focusses on the differences between paper and journal impact. Are there many papers with large differences between both metrics? Is the researcher more able to publish in high-impact journals or papers with high impact? If we use journal percentiles as an expected value, which can be contrasted with paper percentiles, a higher paper percentile demonstrates that the paper received more impact than could be expected on the basis of the publishing journal.

In the scatter and difference against mean plots, which we present in this study, paper and journal impact are categorized into four quadrants. A high general impact of the papers in the set is indicated by high numbers and proportions of data points in quadrants 1 and 4 (i.e., high values of $n_{q1}$ and $n_{q4}$). High numbers and proportions of data points in quadrants 2 and 3 (high values of $n_{q2}$ and $n_{q3}$) indicate a low impact in general. The differences between $n_{q1}$ and $n_{q4}$ as well as $n_{q2}$ and $n_{q3}$ indicate if the papers show a higher journal or higher paper impact. With the categorization of journal and citation impact into four quadrants, the plots follow approaches in bibliometrics, such as the Characteristics Scores and Scales (CSS) method (Glänzel, Debackere, & Thijs, 2016), which can also be used to assign citation impact to (four) impact groups.

By exploring paper impact and journal impact using combined plots, the user should be aware that both perspectives on researchers' performance are frequently related. Many studies have investigated the correlation between journal metrics and citation impact on the single paper level. Overviews of these studies can be found in Bornmann and Daniel (2008), Onodera and Yoshikane (2014), and Tahamtan, Safipour Afshar, and Ahamdzadeh (2016). Most of these studies have revealed that paper and journal impact are positively correlated: one can expect more citations if the manuscript has been published in a high-impact journal and fewer citations if it is a low impact journal. This means for the use of the plots, which have been introduced and explained in this paper, that similar results can be expected on the paper as well as the journal side. Thus, the most interesting parts of the plots are those, which visualize information about papers with disagreeing journal and paper impact.

# Acknowledgements

# References

Altman, D. G., & Bland, J. M. (1983). Measurement in Medicine: The Analysis of Method Comparison Studies. *Journal of the Royal Statistical Society. Series D (The Statistician), 32*(3), 307-317. doi: 10.2307/2987937.

Bornmann, L. (2011). Scientific Peer Review. *Annual Review of Information Science and Technology, 45*, 199-245.

Bornmann, L., & Daniel, H.-D. (2008). What do citation counts measure? A review of studies on citing behavior. *Journal of Documentation, 64*(1), 45-80. doi: 10.1108/00220410810844150.

Bornmann, L., de Moya Anegón, F., & Leydesdorff, L. (2012). The new Excellence Indicator in the World Report of the SCImago Institutions Rankings 2011. *Journal of Informetrics, 6*(2), 333-335. doi: 10.1016/j.joi.2011.11.006.

Bornmann, L., Leydesdorff, L., & Mutz, R. (2013). The use of percentiles and percentile rank classes in the analysis of bibliometric data: opportunities and limits. *Journal of Informetrics, 7*(1), 158-165.

Bornmann, L., & Marx, W. (2014a). Distributions instead of single numbers: percentiles and beam plots for the assessment of single researchers. *Journal of the American Society of Information Science and Technology, 65*(1), 206–208.

Bornmann, L., & Marx, W. (2014b). How to evaluate individual researchers working in the natural and life sciences meaningfully? A proposal of methods based on percentiles of citations. *Scientometrics, 98*(1), 487-509. doi: 10.1007/s11192-013-1161-y.

Bornmann, L., & Marx, W. (2014c). The wisdom of citing scientists. *Journal of the American Society of Information Science and Technology, 65*(6), 1288-1292.

Bornmann, L., Mutz, R., Hug, S., & Daniel, H. (2011). A multilevel meta-analysis of studies reporting correlations between the h index and 37 different h index variants. *Journal of Informetrics, 5*(3), 346-359. doi: 10.1016/j.joi.2011.01.006.

Cleveland, W. S. (1985). *The elements of graphing data*: Wadsworth Advanced Books and Software.

Garfield, E. (2006). The history and meaning of the Journal Impact Factor. *Journal of the American Medical Association, 295*(1), 90-93.

Glänzel, W., Debackere, K., & Thijs, B. (2016). Citation classes: a novel indicator base to classify scientific output. Retrieved October, 21, 2016, from https://www.oecd.org/sti/051%20-%20Blue%20Sky%20Biblio%20Submitted.pdf

Haunschild, R., & Bornmann, L. (2016). The proposal of using scaling for calculating field-normalized citation scores. *El Profesional de la información, 25*(1), 1699-2407.

Hazen, A. (1914). Storage to be provided in impounding reservoirs for municipal water supply. *Transactions of American Society of Civil Engineers, 77*, 1539-1640.

Hirsch, J. E. (2005). An index to quantify an individual's scientific research output. *Proceedings of the National Academy of Sciences of the United States of America, 102*(46), 16569-16572. doi: 10.1073/pnas.0507655102.

Lariviere, V., Kiermer, V., MacCallum, C. J., McNutt, M., Patterson, M., Pulverer, B., . . . Curry, S. (2016). A simple proposal for the publication of journal citation distributions. *bioRxiv*. doi: 10.1101/062109.

Leydesdorff, L., Wouters, P., & Bornmann, L. (2016). Professional and citizen bibliometrics: complementarities and ambivalences in the development and use of indicators—a state-of-the-art report. *Scientometrics, 109*(3), 2129–2150. doi: 10.1007/s11192-016-2150-8.

Onodera, N., & Yoshikane, F. (2014). Factors affecting citation rates of research articles. *Journal of the Association for Information Science and Technology, 66*(4), 739–764. doi: 10.1002/asi.23209.

Pudovkin, A. I., & Garfield, E. (2004). Rank-normalized impact factor: a way to compare journal performance across subject categories. In J. B. Bryans (Ed.), *ASIST 2004: Proceedings of the 67th Asis&T Annual Meeting, Vol 41, 2004: Managing and Enhancing Information: Cultures and Conflicts* (Vol. 41, pp. 507-515). Medford: Information Today Inc.

Tahamtan, I., Safipour Afshar, A., & Ahamdzadeh, K. (2016). Factors affecting number of citations: a comprehensive review of the literature. *Scientometrics, 107*(3), 1195-1225. doi: 10.1007/s11192-016-1889-2.